\documentclass[twocolumn]{svjour3}         
\smartqed  
\usepackage{graphicx}
\newcommand{\arcsec}{\hbox{$^{\prime\prime}$}}
\newcommand{\arcmin}{\hbox{$^\prime$}}
\journalname{Astrophysics and Space Science}
\begin{document}

\title{ALMA as the ideal probe of the solar chromosphere}
\titlerunning{ALMA as the ideal probe of the solar chromosphere}        

\author{Maria A. Loukitcheva \and Sami K. Solanki \and Stephen White}
\institute{M.A. Loukitcheva \and S.K. Solanki
          \at Max-Planck-Institut f\"ur Sonnensystemforschung, D-37191
Katlenburg-Lindau, Germany \\
              \email{lukicheva@mps.mpg.de}           
           \and
           M.~A.~Loukitcheva \at Astronomical Institute, St. Petersburg University, 198504 St. Petersburg,
           Russia\\
           \and
           S.~White \at Astronomy Department, University of Maryland, College Park, MD
           20742, USA}

\date{Received: date / Accepted: date}

\maketitle

\begin{abstract}
The very nature of the solar chromosphere, its structuring and
dynamics, remains far from being properly understood, in spite of
intensive research. Here we point out the potential of chromospheric
observations at millimeter wavelengths to resolve this long-standing
problem. Computations carried out with a sophisticated dynamic model
of the solar chromosphere due to Carlsson and Stein demonstrate that
millimeter emission is extremely sensitive to dynamic processes in
the chromosphere and the appropriate wavelengths to look for dynamic
signatures are in the range 0.8-5.0 mm. The model also suggests that
high resolution observations at mm wavelengths, as will be provided
by ALMA, will have the unique property of reacting to both the hot
and the cool gas, and thus will have the potential of distinguishing
between rival models of the solar atmosphere. Thus, initial results
obtained from the observations of the quiet Sun at 3.5 mm with the
BIMA array (resolution of 12\arcsec) reveal significant oscillations
with amplitudes of 50-150 K and frequencies of 1.5-8 mHz with a
tendency toward short-period oscillations in internetwork and longer
periods in network regions. However higher spatial resolution, such
as that provided by ALMA, is required for a clean separation between
the features within the solar atmosphere and for an adequate
comparison with the output of the comprehensive dynamic simulations.

\end{abstract}

\section{Introduction}
\label{intro}

The chromosphere remains the least understood layer of the solar
atmosphere, with the very basics of its structure being hotly
debated: is it better described by the classical picture of a steady
temperature rise as a function of height, with superposed weak
oscillations (e.g. semi empirical models of Vernazza et al.
\cite{vernazza}, Fontenla et al. \cite{fontenla}), or does the
temperature keep dropping outwards, with very hot shocks producing
strong localized heating (radiation hydrodynamic simulations of
Carlsson \& Stein \cite{CS95}, \cite{CS02}, and Wedemeyer et al.
\cite{W04a})? The latter concept is consistent with the IR
observations of carbon monoxide, which require cool gas to be
present at chromospheric heights (see, e.g. Ayres \cite{ayres}).

\begin{figure*}
\begin{centering}
\includegraphics[width=0.75\textwidth, angle=90]{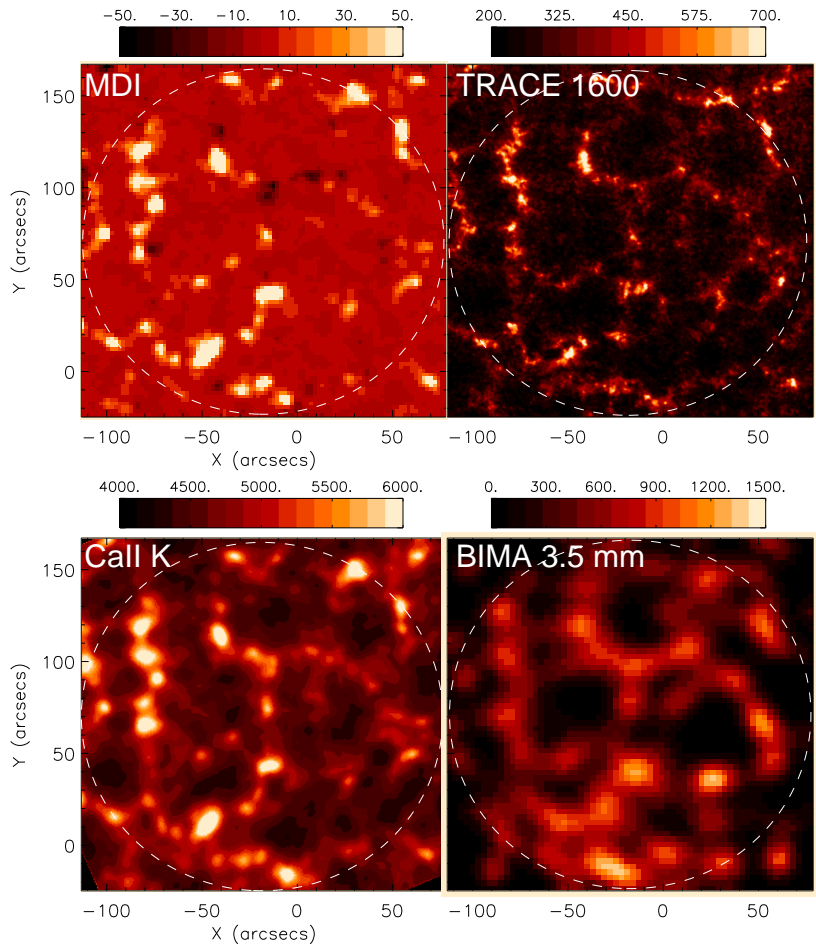}
\caption{Portrait of the solar chromosphere at the center of the
Sun's disk at 4 different wavelengths on May 18, 2004. From top left
to bottom right:
 MDI longitudinal photospheric magnetogram,
 UV 1600 A image from TRACE, CaII K line center image from BBSO and
 BIMA image at 3.5 mm. }
\label{fig:1}       
\end{centering}
\end{figure*}
Thus, existing models cannot provide a complete description of the
solar chromosphere. Consequently now-\\adays two alternative
pictures of the chromosphere co-exist and the role played by
chromospheric dynamics in the structuring of this atmospheric layer
is a subject of intense scientific debate.

One reason for conflicting models is that they are based either on
atomic chromospheric lines and continua in the UV or on molecular
lines in the IR, since UV observations are practically blind to cool
gas in a dynamic chromosphere, while the IR observations sample only
the cool part of the chromosphere. Improved and more sensitive
diagnostics of the chromospheric structure and dynamics, that sample
both the hot and the cool gas and should distinguish between the
rival models, are provided by observations at millimeter wavelengths
with an acceptable spatial resolution as was proposed by Loukitcheva
et al. \cite{L04}. In this contribution we review the unique
chromospheric observations at $3.5$~mm with the
Berkeley-Illinois-Maryland Array and the analysis of the intensity
variations expected from the model of Carlsson \& Stein for mm
wavelengths. We postulate the requirements for mm observations with
the future instruments, with emphasis on spatial and temporal
resolution. Finally we discuss the prospects for chromospheric
studies with ALMA. \vskip-2cm
\section{Results}
\subsection{Analysis of the BIMA observations at 3.5 mm}
\label{sec:1}

The Berkeley-Illinois-Maryland Array (BIMA) operating at a
wavelength of 3.5~mm (frequency of 85~GHz) has been the only
interferometer in the mm range frequently used for solar
observations. The BIMA telescopes are now part of the CARMA array
which will also carry out such observations. With the BIMA data
obtained in the years 2003 and 2004 we have constructed
two-dimensional maps of the solar chromosphere with a resolution of
12\arcsec, which represents the highest spatial resolution achieved
so far at this wavelength for non-flare solar observations. The BIMA
images have led to new insights in to chromospheric structure and to
the detection of spatially-resolved chromospheric oscillations at mm
wavelengths. The details of the restoration procedure and extensive
tests of the sensitivity of the BIMA data to the detection of
dynamic signatures can be found in White et al. \cite{White06}.

 With the currently available resolution the contrast of the brightness
 structures is evaluated to be up to 30\% of the quiet-sun brightness
 at 3.5~mm (White et al. \cite{White06}).
 However, the similarity of brightness structures, derived from the mm images and seen in other
 chromospheric emissions (Fig.\ref{fig:1}), in spite of the difference in resolution of the images
 (1-2\arcsec\ resolution of the UV images), implies that
 the BIMA resolution is not enough to resolve the millimeter fine structure
 and observations with spatial resolution much higher than 12\arcsec\ are required.
 A detailed analysis of the relations between the millimeter emission,
 magnetic field and other chromospheric diagnostics
 is in preparation.

In the millimeter brightness we detected intensity oscillations with
typical amplitudes of 50-150 K in the range of periods from 120 to
700 seconds (frequency range 1.5-8 mHz). We found a tendency toward
short period oscillations in internetwork and longer periods in
network regions in the quiet Sun, which is in good agreement with
the results obtained at other wavelengths. At 3 mm the inner parts
of the chromospheric cells exhibit a behavior typical of the
internetwork with the maximum of the Fourier power in the 3-minute
range, however, most of the oscillations are quasi-periodic, showing
up in wave trains of finite duration lasting for typically 1-3 wave
periods (see also Loukitcheva et al. \cite{L06}).

\subsection{Analysis of the CS model millimeter spectrum}
\label{sec:2}

\begin{figure}[!h]
\begin{centering}
  \includegraphics[width=0.53\textwidth]{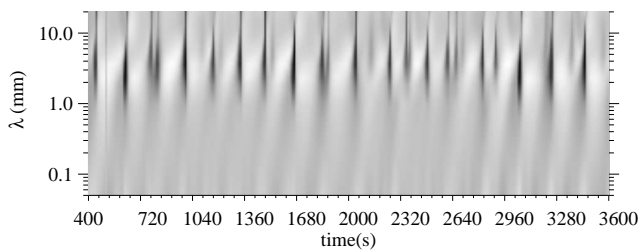}
\caption{Evolution of the Carlsson \& Stein model millimeter
spectrum with time. Negative grey scale representing excess
intensity as a function of time and wavelength.}
\label{fig:2}       
\end{centering}
\end{figure}

The response of the submillimeter and millimeter radiation to a
time-series generated by Carlsson \& Stein (CS) was computed under
the assumption of thermal free-free radiation by Loukitcheva et al.
\cite{L04}. The results are depicted in Fig.~\ref{fig:2} as the
excess intensity as a function of wavelength and time. \vskip-1.8cm
 \vskip-0.5cm
Wave periods of approximately 3 min can be clearly distinguished in
the intensity at all considered wavelengths. Though the dominant
frequency of the oscillations changes slightly with wavelength, for
all mm wavelengths it lies in the range of 3 minutes. The difference
from one period of time to another can be explained by the presence
of merging shocks during certain time intervals. The differences in
the light curves at different wavelengths are caused primarily by
the difference in the formation heights of the emitted radiation. In
general the amplitudes of the oscillations compared to the radiation
temperature are large, in this sense mm wavelength radiation
combines the advantages of the CO lines, which mainly see the cool
gas, with those of atomic lines and UV continua, which mainly sample
the hot gas.

On the whole, the brightness temperatures are extremely
time-dependent at millimeter wavelengths, following changes in the
atmospheric parameters. With increasing wavelength the amplitude of
the brightness oscillations grows significantly, reaches its maximum
value at 2.2 mm (expected to be 15\% of the quiet-Sun brightness
temperature), and decreases rapidly towards longer wavelengths. Thus
we can identify the range 0.8-5.0 mm as the appropriate range of mm
wavelengths at which one can expect the clearest signatures of
dynamic effects. A careful look at the mm brightness spectrum as a
function of time (see Fig.~\ref{fig:2}) reveals a time delay between
the oscillations at long and short millimeter wavelengths. Hence, it
is possible to study wave modes traveling in the chromosphere by
comparing sub-mm with mm observations.

\section{Discussion}
 The CS model predicts that spatially and temporally resolved observations should clearly exhibit
 the signatures of the strong shock waves. However,
 a direct comparison of the observational data products (RMS
values, histogram skewness, Fourier and wavelet spectra, etc.),
referring to regions with weak magnetic field like the quiet Sun
internetwork, with the corresponding products expected from the
simulations of Carlsson \& Stein exhibits large differences. In
particular, the RMS of the brightness temperature is nearly an order
of magnitude larger in the model (800~K at 3~mm) than in the
observations (100~K). Another difference is the absence of longer
periods in the model power spectrum. But these discrepancies do not
rule out the CS models. On the one hand the model is one dimensional
and hence does not predict a coherence length of the oscillations,
while on the other hand we are not able to resolve individual oscillating
elements due to the limited spatial resolution of the observations.

Consequently we estimated the influence of the spatial smearing on
the model parameters of chromospheric dynamics and on the observed
oscillatory power. Thus we confirmed that the very limited spatial
resolution currently available hinders a clean separation between
cells and network and typically both network and internetwork areas
contribute to the recorded BIMA radiation. From the analysis of the
observational data it was found that power in all frequency ranges
increases significantly with improving resolution. Consistency
between the power predicted by the CS model and the observed power
is obtained if the coherence length of oscillating elements is on
the order of 1\arcsec.

Our results are consistent with Wedemeyer et al. \cite{W04b}, who
computed the millimeter wave signature resulting from the 3-D
simulations of Wedemeyer et al. \cite{W04a}. Although the 3-D
simulations suffer from the fact that the radiative transfer of
energy is computed entirely in LTE, which becomes a poor assumption
at chromospheric heights, the authors believe that the chromospheric
pattern and its temporal evolution is representative of the
non-magnetic internetwork regions of the solar chromosphere. The
simulations display a complex 3D structure of the chromospheric
layers, which is highly dynamical on temporal scales of 20-25 s and
on spatial scales comparable to solar granulation, which is in good
agreement with the 1\arcsec\ size of oscillating elements that we
deduced. According to Wedemeyer et al. \cite{W04a} the chromospheric
temperature structure is characterized by a pattern of hot shock
waves, which originate from convective motions, and cool gas lying
between the shocks. The intensity distribution at mm wavelengths
follows the pattern of the shocks in the chromosphere with a sub
arcsecond size of the features associated with the shocks. All this
complex and dynamic 3D structure can be deduced from observations at
mm wavelengths with a sufficiently high spatial resolution of better
than 1\arcsec.

\section{Summary}

Simultaneous mm-submm observations at different wavelengths can be
used for the tomography of the solar atmosphere, as radiation at the
different wavelengths originates from different layers, with the
average formation height increasing with wavelength. Such
observations also provide a strong test of present and future
models. However, observations that might be able to uncover the
nature of the chromosphere should meet the following requirements:
\begin{itemize}
    \item multiband observations in mm-submm domain (0.8-5.0~mm) to address shock waves
and chromospheric oscillation modes
    \item arcsecond spatial resolution to resolve fine structure
    \item temporal resolution better than a few seconds to follow its evolution in time
    \item FOV size of order of 1\arcmin\ 
\item accurate absolute calibration of the observations (Bastian \cite{bastian})
\end{itemize}

These requirements look very similar to the technical specification
of the continuum observations with the Atacama Large Millimeter
Array (ALMA), which represents an enormous advance over existing
instrumentation operating at mm-submm wavelengths. ALMA will produce
images of the highest resolution available for the foreseeable
future (although the technical problem of sampling both large and
small spatial scales simultaneously, required for high--quality imaging
of the chromosphere, will remain a challenge)
and will be the most sensitive instrument operating at submm-mm wavelengths.
To summarize, ALMA will be an extraordinarily
powerful instrument for studying the solar chromosphere. It will
finally allow the mapping of the three-dimensional thermal structure
of the solar chromosphere which will be a real breakthrough in solar
studies.

\begin{acknowledgements}
The use of BIMA for scientific research carried out at the
University of Maryland is supported by NSF grant AST--0028963. Solar
research at the University of Maryland is supported by NSF grant ATM
99-90809 and NASA grants NAG 5-8192, NAG 5-10175, NAG 5-12860 and
NAG 5-11872.
\end{acknowledgements}


\end{document}